\documentclass[superscriptaddress,pre,aps,twocolumn,showpacs,floatfix]{revtex4-1}

\usepackage[dvips]{graphicx}
\usepackage{amsmath}
\usepackage{amssymb}
\usepackage[nice]{nicefrac}
\usepackage{color}

\begin{document}
\bibliographystyle{prsty}
\title{Fluctuation-dissipation relations under L\'evy noises}

\author{Bart{\l}omiej Dybiec}
\email{bartek@th.if.uj.edu.pl}
\affiliation{Center for Models of Life, Niels Bohr Institute, University of Copenhagen, Blegdamsvej 17, 2100 Copenhagen \O, Denmark}
\affiliation{Marian Smoluchowski Institute of Physics, and Mark Kac Center for Complex Systems Research, Jagellonian University, ul. Reymonta 4, 30--059 Krak\'ow, Poland }

\author{Juan M. R. Parrondo}
\email{parrondo@fis.es}
\affiliation{Departamento de F\'isica At\'omica, Molecular y Nuclear and GISC, Universidad Complutensede Madrid, 28040-Madrid, Spain}

 \author{Ewa Gudowska-Nowak}
\email{gudowska@th.if.uj.edu.pl}
 \affiliation{Marian Smoluchowski Institute of Physics, and Mark Kac Center for Complex Systems Research, Jagellonian University, ul. Reymonta 4, 30--059 Krak\'ow, Poland}

\date{\today}
\begin{abstract}
For systems close to equilibrium, the relaxation properties of
measurable physical quantities are described by the linear response
theory and the fluctuation-dissipation theorem (FDT). Accordingly,
the response or the generalized susceptibility, which is a function
of the unperturbed equilibrium  system, can be related to the
correlation between spontaneous fluctuations of a given {\em
conjugate} variable. There have been several attempts to extend the
FDT far from equilibrium, introducing new terms or using effective
temperatures. Recently, Prost, Joanny, and Parrondo [Phys. Rev.
Lett. {\bf 103}, 090601 (2009)] have shown that the FDT can be
restored far from equilibrium by choosing the appropriate variables
conjugate to the external perturbations. Here, we apply the
generalized FDT to a system perturbed by time-dependent
deterministic forces and acting under the influence of white
$\alpha$-stable noises.
\end{abstract}

\maketitle


\section{Introduction\label{sec:introduction}}

The fluctuation-dissipation theorem (FDT)  connects correlation
functions to linear response functions and constitutes a useful tool
in investigations of physical properties of systems at thermodynamic
equilibrium \cite{kubo1966,toda1991}. By virtue of the FDT,
measurable macroscopic physical quantities like  specific heats,
susceptibilities or compressibilities can be related to correlation
functions of spontaneous fluctuations. For systems weakly displaced
from equilibrium, the FDT allows one to express the linear response of
physical observables to time-dependent external fields in terms of
time-dependent correlation functions. Accordingly, departures from
the FDT can be expected for far-from equilibrium situations and have
been demonstrated in various aging, glassy and biological media
\cite{hayashi2007,crisanti2003,calabrese2005,marconi2008}.

On the other hand, the wealth of  theoretical, experimental and
numerical research indicate that the FDT is a special case  of more
general fluctuation relations that remain valid also in a specific
class of nonequilibrium systems
\cite{chetrite2008,speck2010,touchette2007,touchette2009,marconi2008,chechkin2009,dubkov2010,allegrini2009,ebeling2005}.
Following a former generalization of FDT \cite{prost2009} based on the
identity derived by Hatano and Sasa \cite{hatano2001}, we discuss
here an extension of the fluctuation theorem to stochastic models
obeying  Markovian dynamics and driven by white $\alpha$-stable noises. We
apply the generalized fluctuation-response theorem to this case and
analyze the regime in which linear response theory becomes
invalidated. We illustrate our results with the simple example of an
oscillator coupled to a non-equilibrium bath whose action is
represented by a L\'evy white noise.

Let us review the FDT introduced in \cite{prost2009} and extended to arbitrary observables in \cite{speck2010}. The theorem
applies to any Markov process $x(t)$ whose dynamics depends on a
set of parameters $\vec\lambda$. We study the linear response of the
system to perturbations
$\vec\lambda(t)=\vec\lambda_0+\delta\vec\lambda(t)$ around a
reference state, $\rho_{\rm ss}(x;\vec\lambda_0)$ which is the stationary state corresponding to
constant parameters $\vec\lambda_0$. Given an arbitrary observable
$A(x)$, the response can be written as
\begin{equation}
\langle A(t)\rangle-\langle A\rangle_0 \simeq \int_0^t
\chi_{A,\gamma}(t-t')\delta\lambda_\gamma(t')dt',
\label{fdr0}
\end{equation}
where $A(t)\equiv A(x(t))$ and the brackets $\langle\dots\rangle_0$ indicate an average over
the reference state $\rho_{\rm ss}(x;\vec\lambda_0)$; summation over the repeated index $\gamma$ is assumed and
$\chi_{A,\gamma}$ is the time-dependent susceptibility of variable
$A$ with respect to variations of $\lambda_{\gamma}$ (i.e. perturbations of the $\gamma$ component of $\vec\lambda$). The FDT
relates this susceptibility to correlations measured in the
reference unperturbed state
\begin{equation}
\chi_{A,\gamma}(t-t')=\frac{d}{dt}\langle A(t)X_\gamma(t')\rangle_0,
\label{eq:fdt}
\end{equation}
where  $X_\gamma(x)$ is the variable {\em
conjugate} to the perturbation $\lambda_\gamma$ which is defined as
\begin{equation}
X_\gamma(x) = \left.-\frac{\partial \ln \rho_{\rm ss}(x;\vec\lambda)}
{\partial\lambda_\gamma}\right|_{\vec\lambda=\vec\lambda_0}.
\label{conjugate}
\end{equation}
In this definition $\rho_{\rm ss}(x;\vec\lambda)$ is the stationary
probability distribution function (PDF) for constant values of the parameters
$\vec\lambda$.

If the reference state is the Gibbs equilibrium state corresponding to a
temperature $kT=\beta^{-1}$ and a  Hamiltonian ${\cal
H}(x;\vec\lambda)$, the stationary PDF $\rho_{\rm ss}(x;\vec\lambda)$ assumes the form
$\rho_{\rm ss}(x;\vec\lambda)=\exp [ {-\beta{\cal
H}(x;\vec\lambda)}]/Z(\beta,\vec\lambda)$
and the conjugate variable reads
\begin{equation}
X_\gamma(x) = \left.\frac{1}{kT}\frac{\partial \left[{\cal H}(x;\vec\lambda)-{ F}(\beta,\vec\lambda)\right]}
{\partial\lambda_\gamma}\right|_{\vec\lambda=\vec\lambda_0},
\end{equation}
where ${ F}=-kT\ln Z$ stands for the free energy.  Accordingly,
$X_\gamma$ can be  interpreted  as the fluctuation of the quantity $\frac{\partial {\cal H}(x;\vec\lambda_0)}
{\partial\lambda_\gamma}\equiv\left.\frac{\partial {\cal H}(x;\vec\lambda)}
{\partial\lambda_\gamma}\right|_{\vec\lambda=\vec\lambda_0}$:
\begin{equation}
X_\gamma(x) = \frac{1}{kT}\left[\frac{\partial {\cal H}(x;\vec\lambda_0)}
{\partial\lambda_\gamma}-\left\langle \frac{\partial {\cal H}(x;\vec\lambda_0)}
{\partial\lambda_\gamma}\right\rangle_0\,\right].
\end{equation}
For instance, if $\lambda_\gamma$ is a force
coupled to a coordinate $x_\gamma$, i.e., if the control parameter
appears in the Hamiltonian as $-\lambda_\gamma x_\gamma$, then the  conjugate variable
$X_\gamma=-(x_\gamma-\langle x_\gamma\rangle)/(kT)$ represents
fluctuations of $x_{\gamma}$.

On the other hand, if the reference state $\rho_{\rm ss}(x;\vec\lambda_0)$ is not an equilibrium state, the conjugate variables defined by Eq.~\eqref{conjugate} do not have any straightforward physical interpretation \cite{prost2009,speck2010}. In this Letter, we apply the generalized FDT to
a system obeying non-equilibrium Markovian dynamics
 driven by L\'evy white noise. The system of this
type may be conceived as a generalization of Brownian motion: the particle
undergoing L\'evy superdiffusion is performing motion with random jumps and step
lengths following a power-law distribution.  As a result, the width of the
distribution of particles grows superlinearly with time
\cite{dubkov2008,dybiec2009h} signaling anomalous dynamics. A
characteristic feature of L\'evy flights is the dominance of rare but large jumps.
This type of anomalous transport has been
found ubiquitous in nature \cite{nielsen2001,dubkov2008} and serves as a
suitable model for various physical phenomena like atmospheric turbulence \cite{shlesinger1986b},
transport in turbulent plasmas \cite{sanchez2008}, activation kinetics in non-equilibrium baths
\cite{dybiec2004}, transport in fractured materials \cite{lee2010},
epidemic spreading \cite{dybiec2008g},
dispersal of banknotes \cite{brockmann2006} or light scattering in heterogeneous dielectric media
\cite{barthelemy2008}.


\section{Linear system driven by L\'evy white noises}

We proceed to discuss response properties of an overdamped L\'evy-Brownian
particle moving in a parabolic potential that is subject to a
deterministic time-dependent force field  $f(t)$ and a white
L\'evy noise $\zeta(t)$ resulting from the fluctuating environment. The corresponding
 Langevin equation reads
\begin{equation}
\left\{
\begin{array}{l}
 \dot{x}(t)=-ax +f(t) + \zeta(t) \\
  x(0)=x_0
\end{array}
\label{eq:langevinsym}
\right..
\end{equation}
The white L\'evy noise $\zeta(t)$ is defined as the time derivative of a stationary L\'evy process \cite{janicki1994,dubkov2008}, i.e., the integral over time
\begin{equation}
L_{\alpha, \beta}(t)\equiv \int_0^t\zeta(s)ds=z(t)
\end{equation}
represents a stochastic process with independent increments  whose probability density $p_{\alpha,\beta}(z,t)$ is a stable L\'evy distribution. Consequently, the  Fourier transform  of the probability density (characteristic function) $\varphi(k,t) = \int_{-\infty}^\infty e^{ikz(t)} p_{\alpha,\beta}(z,t;\sigma_0,\mu) dz$ reads
\begin{equation}
\varphi(k,t) =\exp\left[ ik\mu_0 t -\sigma_0^\alpha|k|^\alpha t\left( 1-i\beta\,\mbox{sign}(k)
\tan\frac{\pi\alpha}{2} \right)\right]
\label{eq:charakt}
\end{equation}
for $\alpha\neq 1$ and
\begin{equation}
\varphi(k,t) = \exp\left[ ik\mu_0 t -\sigma_0|k| t\left( 1+i\beta\frac{2}{\pi}\mbox{sign} (k) \ln|k| \right) \right]
\label{eq:charakt1}
\end{equation}
for $\alpha=1$ \cite{feller1968}. Here $\alpha\in (0, 2]$ is the L\'evy (stability) index,  $\beta \in \left[ - 1, 1 \right]$ is the skewness parameter (for $\beta=0$ the distributions are symmetric), $\sigma_0 > 0$ represents the noise intensity, and $\mu_0\in \mathbb{R}$ is a location (shift) parameter.
The Gaussian
distribution is a special case of symmetric stable distribution with
$\alpha=2$ and $\beta=0$. In this case  $\mu_0t$ is the mean of the Gaussian random variable $z(t)$ and $\sigma_0^2t$ is its variance.
For $\alpha <2$, stable probability densities exhibit long tails and divergent moments: the asymptotic
(large $z$) behavior of the corresponding PDF is then characterized by a power-law  $p_{\alpha,\beta}(z,t;
\sigma_0,\mu) \propto |z|^{-(1+\alpha)}$.
Under those circumstances, equation \eqref{eq:langevinsym} is associated
with the space-fractional Fokker-Planck-Smoluchowski equation (FFPE)
\cite{metzler1999,yanovsky2000,schertzer2001}:
\begin{eqnarray}
\label{eq:ffpe}
& &\frac{\partial p(x,t)}{\partial t}=-\frac{\partial}{\partial x}\left[ \mu_0 - ax+f(t) \right] p(x,t) \\
  & & + \sigma_0^\alpha \frac{\partial^\alpha}{\partial |x|^\alpha} p(x,t)
 +  \sigma_0^\alpha \beta\tan\frac{\pi\alpha}{2}\frac{\partial}{\partial
x} \frac{\partial^{\alpha-1}}{\partial |x|^{\alpha-1}} p(x,t).
 \nonumber
\end{eqnarray}
Here, the fractional (Riesz-Weyl) derivative is defined by its Fourier transform
$\mathcal{F}\left[\frac{\partial^\alpha}{\partial|x|^\alpha}f(x) \right]=-|k|^\alpha \mathcal{F}\left[ f(x)\right]$
\cite{podlubny1998,jespersen1999}.
Accordingly, Eq.~\eqref{eq:ffpe} has the following Fourier representation
\begin{eqnarray}
\label{eq:fourierrepresentationfp}
\frac{\partial \hat{p}(k,t)}{\partial t} & = &  -ak\frac{\partial}{\partial
k} \hat{p}(k,t) +ik \left[\mu_0+f(t)\right]\hat p(k,t)
\\
\nonumber
& & - \sigma_0^\alpha |k|^\alpha\left[1-i\beta\,\mbox{sign}(k)
\tan\frac{\pi\alpha}{2} \right] \hat{p}(k,t),
\end{eqnarray}
where $\hat{p}(k,t)=\mathcal{F} \left[ p(x,t) \right]$.
In what follows, we adhere to the analysis of strictly $\alpha$-stable random variables \cite{janicki1994}, i.e. ones for which $\mu_0=0$, and additionally $\beta=0$ if $\alpha = 1$.

Since our original Langevin equation \eqref{eq:langevinsym} is
linear, its solution depends linearly on the stable
process $L_{\alpha,\beta}(t)$. Accordingly, the probability density of the solution, $p(x,t|x_0,0)$, has
the form of an $(\alpha,\beta)$-stable L\'evy distribution with
time-dependent location $\mu(t)$ and scale $\sigma(t)$ parameters
\cite{jespersen1999}. By analogy, its characteristic function is given by (cf. Eqs.~(\ref{eq:charakt}) and~(\ref{eq:charakt1}))
\begin{equation*}
 \hat p(k,t)=\exp\left[  ik\mu(t)
-\sigma^\alpha(t)|k|^\alpha
\left( 1- i\beta\,
\mathrm{sign}(k)
\tan\frac{\pi\alpha}{2} \right)
 \right].
\end{equation*}
We can now insert this ansatz into the Fokker-Planck-Smoluchowski equation~\eqref{eq:fourierrepresentationfp}. Since the derivative with respect to $k$ appears multiplied by $k$ in \eqref{eq:fourierrepresentationfp}, the non analyticity of $|k|^{\alpha}$ at $k=0$ does not create any singularity in the equation.
The real part of Eq.~\eqref{eq:fourierrepresentationfp}
yields the following evolution equation for the scale parameter $\sigma(t)$
\begin{equation}
 -\alpha\sigma^{\alpha-1}\dot{\sigma}=a\alpha\sigma^\alpha-\sigma_0^\alpha,
\label{eq:sigmaskewed}
\end{equation}
whereas the imaginary part gives
\begin{eqnarray}
 \label{eq:imaginaris}
 \left[ \dot{\mu} + a\mu -f(t) \right]k & =&  \left[
-\alpha\sigma^{\alpha-1}\dot{\sigma}-a\alpha\sigma^\alpha + \sigma_0^\alpha  \right] \\
 & & \times\, \beta\tan\frac{\pi\alpha}{2}|k|^\alpha\,\mathrm{sign}(k) \nonumber.
\end{eqnarray}
The right hand side of Eq.~\eqref{eq:imaginaris} vanishes due to Eq.~\eqref{eq:sigmaskewed}.
From the left hand side one gets the evolution equation for the location parameter $\mu(t)$:
\begin{equation}
 \dot{\mu}(t)=-a\mu+f(t).
 \label{eq:mu}
 \end{equation}
The evolution equations \eqref{eq:sigmaskewed} and \eqref{eq:mu} are completed with the  initial conditions $\mu(0)=x_{0}$ and $\sigma(0)=0$ (we are calculating probability densities conditioned to $x(0)=x_{0}$).  The solution of these differential equations are
\begin{equation}
 \mu(t)=e^{-at}x_0 + e^{-at}\int_0^te^{as}f(s)ds
\label{eq:musol}
\end{equation}
and
\begin{equation}
\sigma(t)=\sigma_0\left[ \frac{1}{a\alpha}\left( 1- e^{-a\alpha t} \right)   \right]^{1/\alpha},
\label{eq:sigmasol}
\end{equation}
where $\sigma_0$ is the scale parameter of the corresponding $\alpha$-stable density.
For a constant force $f(t)\equiv f$, the long time asymptotics
of the above equations are $ \lim_{t\to\infty}\mu(t)=f/a$ and $
\lim_{t\to\infty}\sigma(t)={\sigma_0}/(a\alpha)^{1/\alpha}$.

\section{The conjugate variable}
To determine the conjugate variable to the external force, we need the stationary distribution {\em in position space} for a constant force $f$.  Despite  the characteristic functions  of stable distributions  assume closed expressions, the corresponding PDFs have a known simple analytical form \cite{janicki1994,penson2010} only in a few cases: For $\alpha=2$ and $\beta=0$ the resulting distribution is Gaussian;  for $\alpha=1$, $\beta=0$ one gets the Cauchy distribution; finally, for
 $\alpha=1/2$, $\beta=1$ one gets  the L\'evy-Smirnoff distribution. In this section, we give explicit expressions for the conjugate variable for these three cases.

For $\alpha=2$ and $\beta=0$, the time dependent solution of the corresponding
Langevin Eq.~\eqref{eq:langevinsym} is
\begin{equation}
 p_{2,0}(x,t|x_0,0)=\frac{1}{\sqrt{2\pi\sigma^2(t)}}\exp\left[-\frac{ \left(x-\mu(t)\right)^{2}}{2\sigma^2(t)}\right]
\end{equation}
with  $\mu(t)$ and  $\sigma(t)$ given by Eqs.~\eqref{eq:musol} and~\eqref{eq:sigmasol}. The stationary solution $p_{\rm ss}(x)$ for a constant force $f$ is obtained by replacing $\mu(t)$ and $\sigma^2(t)$ by their stationary values, $f/a$ and $\sigma^2_0/(2a)$, respectively. We then get the
  non-equilibrium potential
\begin{equation}
\phi \equiv -\ln p_{\rm ss}(x)=-\frac{1}{2}\ln\frac{a}{\pi\sigma_0^{2}} + \frac{a(x-f/a)^2}{\sigma_0^2}
\end{equation}
and the conjugate variable can be easily derived as
\begin{equation}
 X_{{\rm G}}=\left.\frac{\partial \phi}{ \partial f}\right|_{f=0}= -\frac{2x}{\sigma_{0}^{2}},
\end{equation}
which is proportional to $x$, as expected, since the Gaussian case corresponds to a Brownian particle in equilibrium.

For $\alpha=1$ and $\beta=0$, the time dependent solution of the corresponding
Langevin Eq.~\eqref{eq:langevinsym} is the Cauchy distribution
\begin{equation}
 p_{1,0}(x,t|x_0,0)=\frac{\sigma(t)}{\pi} \frac{1}{\left[x-\mu(t)\right]^2+\sigma^2(t)}
\label{eq:timedependentcauchy}
\end{equation}
and the stationary solution for a  constant force $f$ is obtained replacing $\mu(t)$ and $\sigma(t)$ by their stationary values, $f/a$ and $\sigma_0/a$, respectively.
The  non-equilibrium potential in this case is given by
\begin{equation}
 \phi =-\ln\frac{\sigma_0}{a \pi} +\ln\left[ {(x-f/a)^2+(\sigma_0/a)^2} \right],
\end{equation}
so that the conjugate variable takes the form
\begin{equation}
 X_{\rm C}= -\frac{2x}{a\left[ x^2+(\sigma_0/a)^2 \right]}
 \label{conjC}
\end{equation}
which is proportional to $x$ only for small values of $x$ and becomes proportional to $1/x$ for large $x$.
This large $x$ behavior ensures the convergence of all the moments of $X_C$, whereas for the Cauchy case $|x|^\nu$ exists only if $\nu<1$, see \cite{janicki1994}

Finally, for $\alpha=1/2$ and $\beta=1$ the solution to Eq.~\eqref{eq:langevinsym} is the L\'evy-Smirnoff PDF
\begin{equation}
 p_{1/2,1}(x,t|x_0,t_0)=\sqrt{\frac{\sigma(t)}{2\pi\left[ x-\mu(t) \right]^{3}}  }
 \exp\left[ -\frac{\sigma(t)}{2(x-\mu(t))} \right]
 \label{eq:levysmirnoff}
\end{equation}
for $x>\mu(t)$ and $ p_{1/2,1}(x,t|x_0,0) \equiv 0$ for $x\leq \mu(t)$. The stationary values of $\mu(t)$ and $\sigma(t)$ are in this case
$f/a$ and $4\sigma_0/a^2$, respectively. Inserting these values, one can easily obtain the  non-equilibrium potential $ \phi=-\ln
p_{\rm ss}(x)$ and the conjugate variable
\begin{equation}
 X_{\rm L-S}=\frac{4\sigma_0-3a^2x}{2a^{3}x^2}; \qquad x>0.
\end{equation}

\section{Susceptibility and response}

The main objective of the current work is to
compare the response of the system to external perturbation
as calculated directly from the definition
\begin{equation}
 \langle X(t) \rangle =   \int_{-\infty}^\infty X(x) p(x,t) dx,
\label{eq:definition}
\end{equation}
or,  otherwise determined by the generalized susceptibility $ \chi(t)=\frac{d}{dt} \langle X(t)X(0) \rangle_{0}$  within linear response theory:
\begin{equation}
 \langle X(t) \rangle_{\rm LR} =   \int_0^t\chi(t-s) f(s) ds.
\label{eq:response}
\end{equation}

We restrict our analysis to the Cauchy case, $\alpha=1,\beta=0$, and denote the conjugate variable as $X\equiv X_{\rm C}$, with $X_{\rm C}$   given by Eq.~(\ref{conjC}).
In this case, the time-dependent average (\ref{eq:definition}) can be calculated exactly with the probability density:
\begin{equation}
 p(x,t)=\int_{-\infty}^\infty p(x,t|x_0,0)p(x_0)dx_0
\end{equation}
where
\begin{equation}
 p(x_0)=\frac{\sigma_0}{a\pi} \frac{1}{x_0^2+(\sigma_0/a)^2}
\label{eq:cauchystationary}
\end{equation}
and $p(x,t|x_0,0)$ is given by Eq.~\eqref{eq:timedependentcauchy}.

On the other hand, the FDT relates the susceptibility with the autocorrelation of the conjugate variables in the reference state, i.e., for $f=0$. The autocorrelation is defined as
\begin{eqnarray}
 \langle X(t)X(0) \rangle_{0} & = &
\iint \frac{2x}{a[x^2+(\sigma_0/a)^2]} \frac{2y}{a[y^2+(\sigma_0/a)^2]} \nonumber \\
& & \;\;\;\;\;\;\times \frac{\sigma(t)}{\pi\left[(x-\mu(t))^2+\sigma^2(t) \right]} \nonumber \\
& & \;\;\;\;\;\;\times \frac{\sigma_0}{a\pi\left[y^2+(\sigma_0/a)^2 \right]} dx dy\nonumber
\end{eqnarray}
where $\mu(t)=e^{-at}y$ and $\sigma(t)=\sigma_0\left[(1-e^{-at})/a \right]$.
The final result is surprisingly simple:
\begin{equation}
  \langle X(t)X(0) \rangle_{0}=  \frac{1}{2\sigma_0^2}e^{-at}.
\end{equation}
From the above, the generalized  susceptibility can be derived by differentiation with respect to time (see Eq.~(\ref{eq:fdt})):
\begin{equation}
 \chi(t)=\frac{d}{dt} \langle X(t)X(0) \rangle_{0}=-\frac{a}{2\sigma_0^2}e^{-at}.
\end{equation}

%
%
\begin{figure}
\begin{center}
\includegraphics[angle=0,width=0.9\columnwidth]{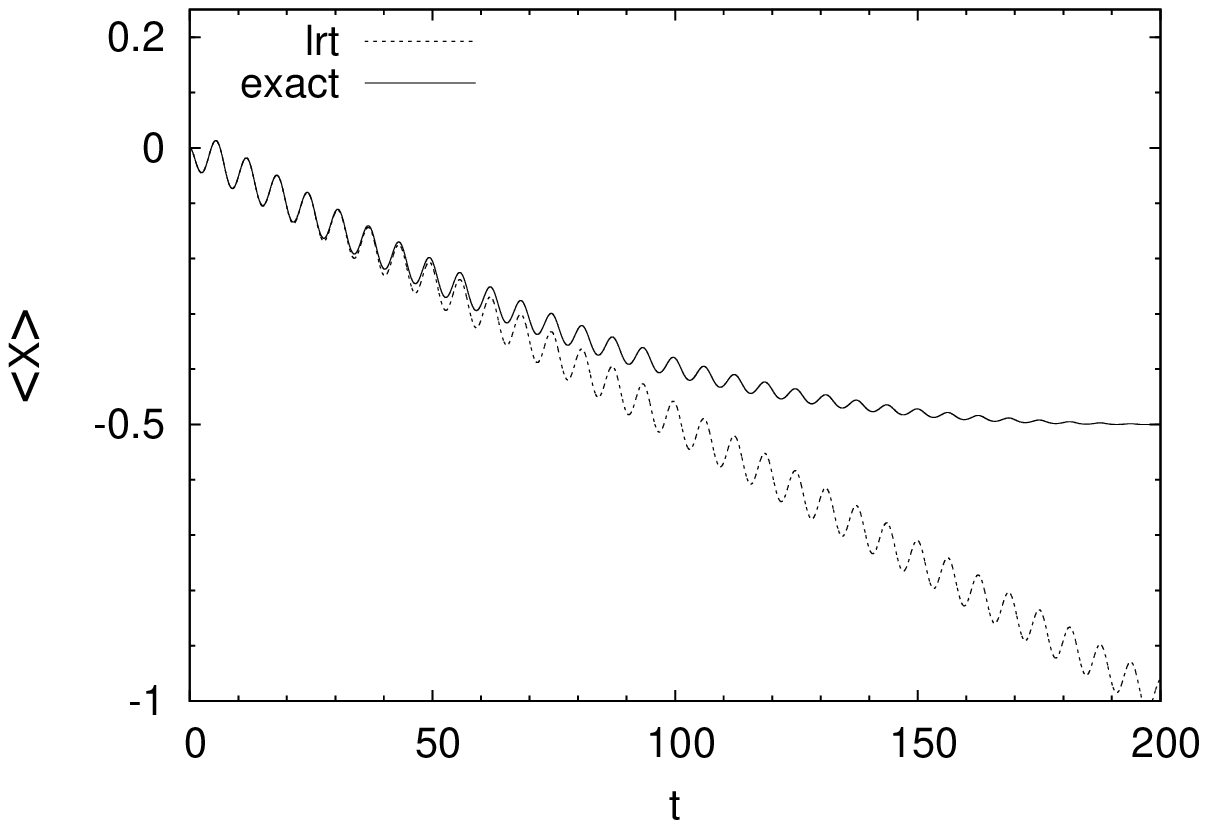}
\includegraphics[angle=0,width=0.9\columnwidth]{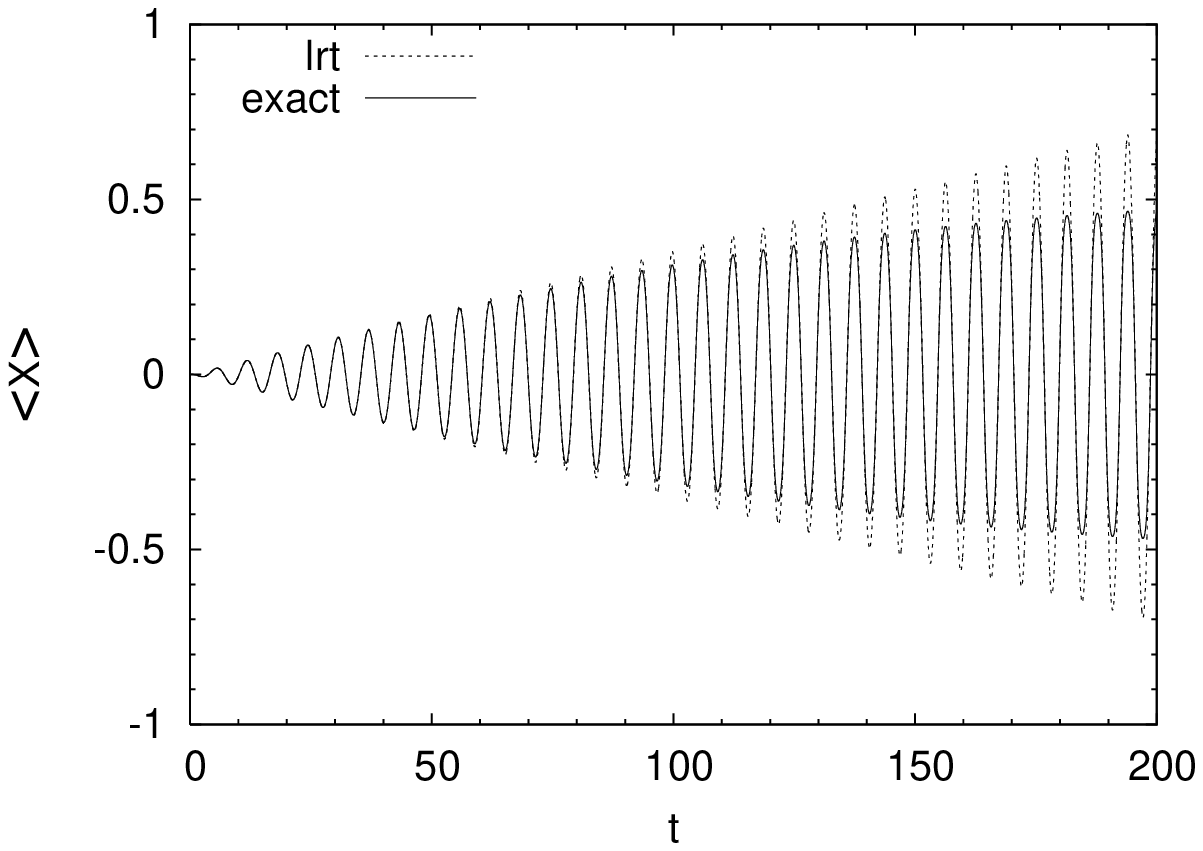}
\caption{Response of $\langle X(t) \rangle$ to external drivings $f_1(t)= {\sin(t)}/{10}+{t}/{100}$ (upper plot) and $f_2(t)=t\sin(t)/100$ (lower plot). The solid and dotted lines present an exact result (Eq.~\eqref{eq:definition}) and a result constructed by use of the linear response theory (Eq.~\eqref{eq:response}), respectively.}
\label{fig:response}
\end{center}
\end{figure}

In further calculations, for the sake of simplicity, it is assumed that $a=1$ and $\sigma_0=1$, so that
$\langle X(t)X(0) \rangle=  \frac{1}{2}e^{-t}$ and $\chi(t) =  -\frac{1}{2}e^{-t}$.

In order to test the linear response theory for our dynamic Markov system subjected to L\'evy white noise, we calculate the response of the conjugate variable $X$ to two different time dependent perturbations: the sum of a small periodic and a linearly increasing force, $f_1(t)=\sin(t)/10+t/100$; and a periodic force with increasing amplitude, $f_2(t)=t\sin(t)/100$. Figure~\ref{fig:response} displays the exact evolution  of $\langle X(t) \rangle$ and the result obtained from the linear response theory. For small perturbations, (i.e. short times) in both cases, linear response theory yields an accurate estimation of the response. In the case of $f_1(t)$, for large times the response $\langle X(t) \rangle$ is insensitive to the sinusoidal component of the force, which is small compared with the linear part.  This is due to the peculiar form of the conjugate variable $X$ given by Eq.~(\ref{conjC}). For a constant force $f$, the mean value of $X$ is
\begin{eqnarray}
\langle X\rangle &=&-\frac{\sigma_0}{a\pi}\int_{-\infty}^{\infty} \frac{dx}{\left[x-f/a\right]^2+(\sigma_0/a)^2}\frac{2x }{a\left[ x^2+(\sigma_0/a)^2 \right]}\nonumber \\
&=& -\frac{2f}{f^2+4\sigma_0^2} \nonumber
\end{eqnarray}
which yields $\langle X\rangle=-0.5$ for $\sigma_0=1$ and $f=2$ (at $t=200$, $f_1(t)\simeq 2$). A similar saturation effect is not observed for the sinusoidal force $f_2(t)$.

\begin{figure}
\begin{center}
\includegraphics[angle=0,width=0.9\columnwidth]{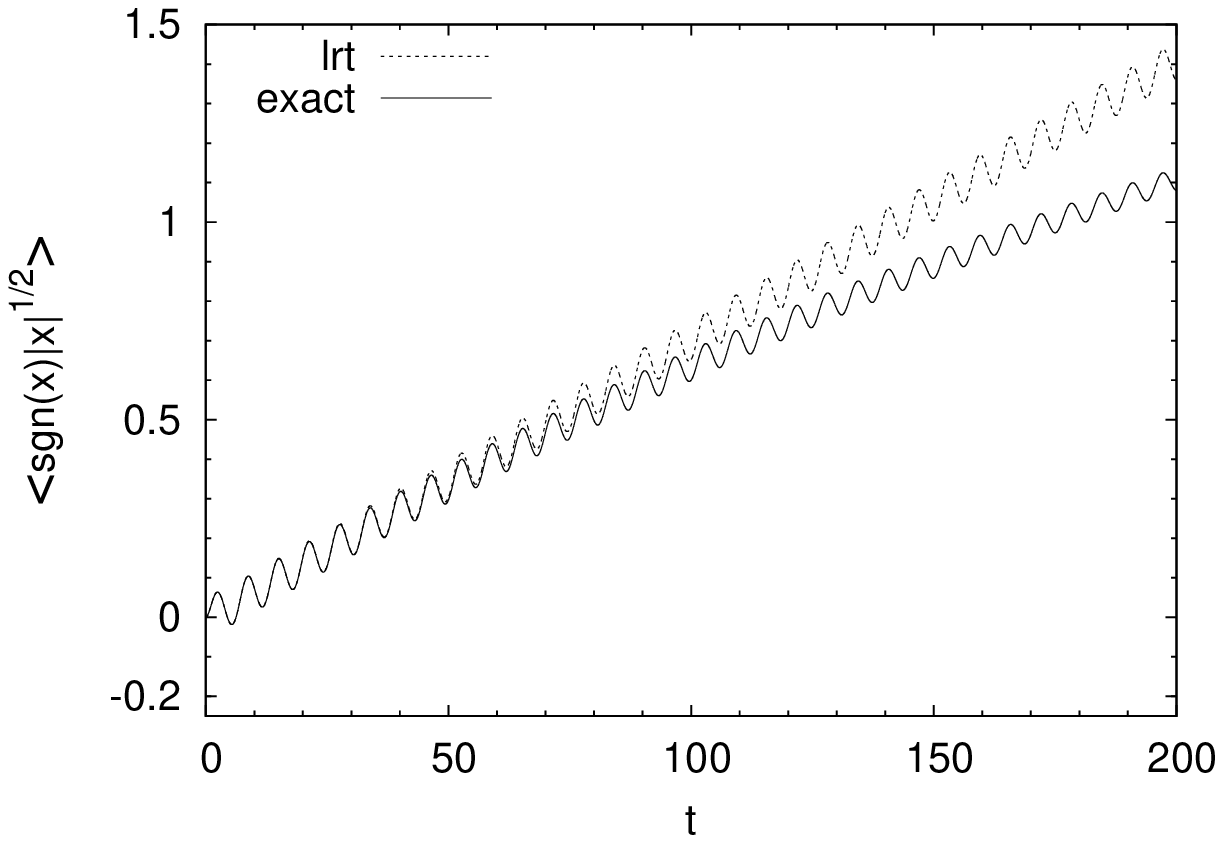}
\includegraphics[angle=0,width=0.9\columnwidth]{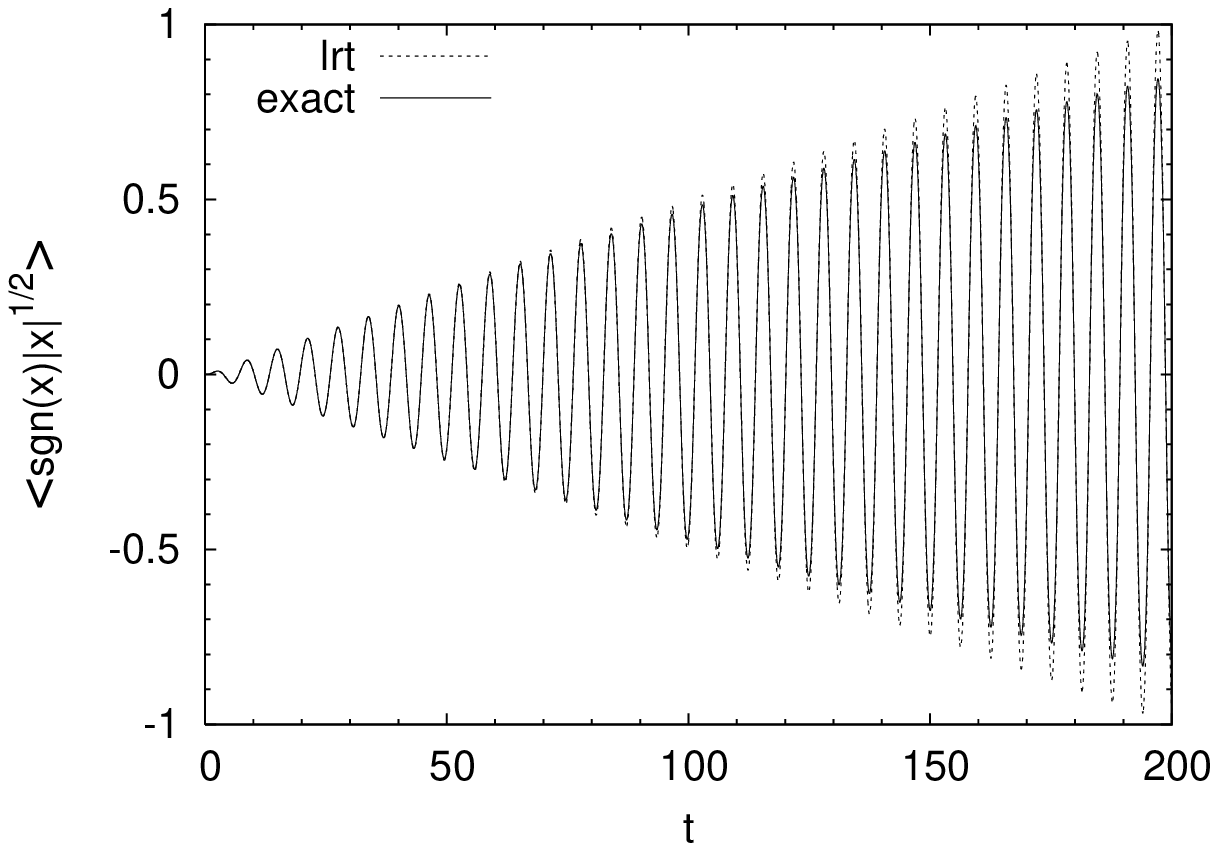}
\caption{Response of the $\langle \mathrm{sign}\left[x(t)\right] | {x(t)} |^{1/2}\rangle$ to $f(t)= {\sin(t)}/{10}+{t}/{100}$ (upper plot) and $f(t)=t\sin (t)/100$ (lower plot).  The solid line presents exact results, see Eq.~(\ref{eq:ssgexact}), while the dotted line results constructed by use of the linear response theory, see Eqs.~\eqref{eq:response} and~\eqref{eq:ssgsusceptibility}.}
\label{fig:sgnmoment}
\end{center}
\end{figure}

We can apply the FDR to any function $A(x)$ with finite average. Due to the long tails of stable distributions, only moments $\langle |x|^\nu \rangle$ with $\nu<\alpha$ converge ($\nu< 1$ in the case of Cauchy distributions) \cite{janicki1994}. Moreover, those moments are even functions of $x$ and, for symmetry reasons, the correlation with $X$ vanishes: $\langle |x(t)|^\nu X(0)\rangle=0$. Consequently, the deviation of  $\langle |x(t)|^\nu \rangle$  with respect to its reference value is non linear in the perturbation $f(t)$. On the other hand, we can obtain non-trivial results for odd fractional moments $
 A(x)=\mathrm{sign} \left[ x(t) \right] |x(t)|^\nu$, whose average in the reference state vanishes $\langle A\rangle_0=0$. The corresponding correlation function reads
\begin{equation}
  \langle A(x(t))X(0) \rangle=  -\frac{\nu}{\sin\left[  \pi\nu/2 \right]}e^{-t}
\label{eq:ssgsusceptibility}
\end{equation}
and the generalized susceptibility is given by
\begin{equation}
  \chi_A(t) =  \frac{\nu}{\sin\left[  \pi\nu/2 \right]}e^{-t}.
\end{equation}
In the spirit of the former definition, see~Eq.~(\ref{eq:definition}), the exact value of $\langle A(x(t)) \rangle$ can be calculated as
\begin{eqnarray}
\label{eq:ssgexact}
 \langle A(x(t)) \rangle& =  & \int_{-\infty}^\infty A(x(t)) p(x,t)dx \\
& = & \int_{-\infty}^\infty A(x(t)) p(x,t|x_0,0)p(x_0)dx, \nonumber
\end{eqnarray}
where $p(x_0)$ and $p(x,t|x_0,0)$ are given by Eqs.~\eqref{eq:timedependentcauchy} and~\eqref{eq:cauchystationary} respectively.
Figure~\ref{fig:sgnmoment} displays the comparison of the exact evolution $\langle \mathrm{sign} \left[ x(t) \right] | {x(t)} |^{1/2} \rangle$  with the linear response approximation $\int_0^t\chi_A(t-s)f(s)ds$. As previously, linear response theory is valid for weak perturbation up to $f\simeq 0.5$. However, for the fractional moment and the linearly increasing force (upper plot), we do not observe saturation.

%
%

\section{Summary and conclusions\label{sec:summary}}
We have shown that the generalized FDT can be applied to linear systems driven by L\'evy noise. The FDT allows one to calculate the susceptibility of any observable and then the response to any small time dependent perturbation. For a noise distributed according to the Cauchy distribution, we have calculated the susceptibility of the conjugate variable $X_{\rm C}$ and the susceptibility of odd fractional moments $\langle {\rm sign} [x(t)] |x(t)^\nu|\rangle$, which have a simple exponential behavior. From these susceptibilities it is easy to get simple analytical expressions for the response of the system using
Eq.~\eqref{eq:response}. We have to notice that, although the exact response can be calculated analytically using Eq.~\eqref{eq:definition}, the corresponding integrals are cumbersome and can be only solved numerically in the simplest cases.
Therefore, the generalized FDT is shown to be a useful analytical tool to deal with these type of systems.

It is still not obvious whether the  conjugate variables that we have calculated for the Cauchy and L\'evy-Smirnoff noises have any physical meaning, besides the one provided by the generalized FDT itself. The generalized FDT shows that these conjugate variables represent the change in the probability distribution of the system under the perturbation. In equilibrium, this change is also related with the energy that the system absorbs from the perturbation. On the other  hand, for non-equilibrium systems, the lack of conserved quantities prevents such an interpretation. For instance, in the case of the harmonic oscillator driven by a Cauchy-L\'evy noise, $\langle x\rangle$ and higher moments diverge (cf.~\cite{footnote}). Consequently, both the potential energy of the system in the harmonic potential and the work done by the external force $f(t)$ also diverge. The system is plagued by divergent quantities. However, the conjugate variable $X_{\rm C}$ given by Eq.~\eqref{conjC} has finite moments and still captures the dynamical response of the L\'evy particle. Summarizing, although systems driven by $\alpha$-stable noises might significantly differ from their Brownian (equilibrium) counterparts \cite{chechkin2002,dybiec2009,dybiec2010d} due to their heavy tail asymptotics,
 we have shown that  in such far-from-equilibrium situations some concepts from weakly perturbed equilibrium systems can be still used.

One of the drawbacks of the generalized FDT derived in \cite{prost2009} is the difficulty to find the conjugate variable, since it requires the knowledge of the stationary state. We have been able to find this stationary state for a linear system. An interesting open question is whether this state, or some slight modification, can still be used to calculate susceptibilities in the presence of weak non-linearities.


\acknowledgments
We are grateful to Jordan Horowitz for reading  the manuscript and providing various useful suggestions.
The authors acknowledge the support by the European Science Foundation (EFS) through Exploring Physics of Small Devices (EPSD) program.
JMRP also acknowledges financial support from grants MOSAICO (Spanish Government) and MODELICO (Comunidad de Madrid).
BD acknowledges the Danish National Research Foundation for financial support through the Center for Models of Life (CMOL).


\begin{thebibliography}{10}
\bibitem{footnote}
With a more general definition of integrals, i.e. by using a ``principal value'' integral, the first moment of a Cauchy distribution can be evaluated as $\langle x\rangle=\lim_{a\rightarrow \infty}\frac{1}{\pi}\int^a_{-a}\frac{x}{1+x^2}dx$ and is equal zero
\end{thebibliography}

\end{document}